\documentclass[aps,prd,twocolumn,showpacs,preprintnumbers,superscriptaddress,amsmath,amssymb]{revtex4}

\usepackage{graphicx}
\usepackage{xspace}
\usepackage{epsfig}
\usepackage{multirow}
\usepackage{dcolumn}  
\usepackage{wasysym}
\usepackage[section]{placeins}

\newcommand{\mb}{\ensuremath{M_{\mathrm{bc}}}\xspace}
\newcommand{\de}{\ensuremath{\Delta E}\xspace}

\newcommand{\br}{\ensuremath{\mathcal{B}}\xspace}

\newcommand{\ks}{\ensuremath{K^0_S}\xspace}
\newcommand{\bb}{\ensuremath{B \overline{B}}\xspace}

\newcommand{\ebeam}{E_\mathrm{beam}}
\newcommand{\btag}{{B_\mathrm{tag}}}

\newcommand{\ecl}{E_\mathrm{ECL}}
\newcommand{\cosb}{\cos\theta_B}
\newcommand{\cost}{\cos\theta_T}

\def\myspecial#1{}                   
\def\calL{{\mathcal L}}

\def\calP{{\mathcal P}}

\def\figurebox#1#2#3{%
    \def\arg{#3}%
    \ifx\arg\empty
    {\hfill\vbox{\hsize#2\hrule\hbox to #2{\vrule\hfill\vbox to #1{\hsize#2\vfill}\vrule}\hrule}\hfill}%
    \else
    {\hfill\epsfbox{#3}\hfill}%
    \fi}

\begin{document}


\myspecial{!userdict begin /bop-hook{gsave 300 50 translate 5 rotate
    /Times-Roman findfont 18 scalefont setfont
    0 0 moveto 0.70 setgray
    (\mySpecialText)
    show grestore}def end}

\title{\quad\\[0.45cm]{\Large \bf \boldmath Search for $B^0$ Decays to Invisible Final States at Belle}}

\affiliation{University of Bonn, Bonn}
\affiliation{Budker Institute of Nuclear Physics SB RAS and Novosibirsk State University, Novosibirsk 630090}
\affiliation{Faculty of Mathematics and Physics, Charles University, Prague}
\affiliation{University of Cincinnati, Cincinnati, Ohio 45221}
\affiliation{Department of Physics, Fu Jen Catholic University, Taipei}
\affiliation{Gifu University, Gifu}
\affiliation{Hanyang University, Seoul}
\affiliation{University of Hawaii, Honolulu, Hawaii 96822}
\affiliation{High Energy Accelerator Research Organization (KEK), Tsukuba}
\affiliation{Indian Institute of Technology Guwahati, Guwahati}
\affiliation{Indian Institute of Technology Madras, Madras}
\affiliation{Institute of High Energy Physics, Chinese Academy of Sciences, Beijing}
\affiliation{Institute of High Energy Physics, Vienna}
\affiliation{Institute of High Energy Physics, Protvino}
\affiliation{Institute for Theoretical and Experimental Physics, Moscow}
\affiliation{J. Stefan Institute, Ljubljana}
\affiliation{Kanagawa University, Yokohama}
\affiliation{Institut f\"ur Experimentelle Kernphysik, Karlsruher Institut f\"ur Technologie, Karlsruhe}
\affiliation{Korea Institute of Science and Technology Information, Daejeon}
\affiliation{Korea University, Seoul}
\affiliation{Kyungpook National University, Taegu}
\affiliation{\'Ecole Polytechnique F\'ed\'erale de Lausanne (EPFL), Lausanne}
\affiliation{Luther College, Decorah, Iowa 52101}
\affiliation{University of Maribor, Maribor}
\affiliation{Max-Planck-Institut f\"ur Physik, M\"unchen}
\affiliation{University of Melbourne, School of Physics, Victoria 3010}
\affiliation{Graduate School of Science, Nagoya University, Nagoya}
\affiliation{Kobayashi-Maskawa Institute, Nagoya University, Nagoya}
\affiliation{Nara Women's University, Nara}
\affiliation{National Central University, Chung-li}
\affiliation{Department of Physics, National Taiwan University, Taipei}
\affiliation{H. Niewodniczanski Institute of Nuclear Physics, Krakow}
\affiliation{Nippon Dental University, Niigata}
\affiliation{Niigata University, Niigata}
\affiliation{University of Nova Gorica, Nova Gorica}
\affiliation{Osaka City University, Osaka}
\affiliation{Pacific Northwest National Laboratory, Richland, Washington 99352}
\affiliation{Panjab University, Chandigarh}
\affiliation{Research Center for Electron Photon Science, Tohoku University, Sendai}
\affiliation{University of Science and Technology of China, Hefei}
\affiliation{Seoul National University, Seoul}
\affiliation{Sungkyunkwan University, Suwon}
\affiliation{School of Physics, University of Sydney, Sydney, New South Wales 2006}
\affiliation{Tata Institute of Fundamental Research, Mumbai}
\affiliation{Excellence Cluster Universe, Technische Universit\"at M\"unchen, Garching}
\affiliation{Tohoku Gakuin University, Tagajo}
\affiliation{Tohoku University, Sendai}
\affiliation{Department of Physics, University of Tokyo, Tokyo}
\affiliation{Tokyo Institute of Technology, Tokyo}
\affiliation{Tokyo Metropolitan University, Tokyo}
\affiliation{Tokyo University of Agriculture and Technology, Tokyo}
\affiliation{CNP, Virginia Polytechnic Institute and State University, Blacksburg, Virginia 24061}
\affiliation{Wayne State University, Detroit, Michigan 48202}
\affiliation{Yamagata University, Yamagata}
\affiliation{Yonsei University, Seoul}

  \author{C.-L.~Hsu}\affiliation{Department of Physics, National Taiwan University, Taipei} 
  \author{P.~Chang}\affiliation{Department of Physics, National Taiwan University, Taipei} 
  \author{I.~Adachi}\affiliation{High Energy Accelerator Research Organization (KEK), Tsukuba} 
  \author{H.~Aihara}\affiliation{Department of Physics, University of Tokyo, Tokyo} 
  \author{K.~Arinstein}\affiliation{Budker Institute of Nuclear Physics SB RAS and Novosibirsk State University, Novosibirsk 630090} 
  \author{D.~M.~Asner}\affiliation{Pacific Northwest National Laboratory, Richland, Washington 99352} 
  \author{V.~Aulchenko}\affiliation{Budker Institute of Nuclear Physics SB RAS and Novosibirsk State University, Novosibirsk 630090} 
  \author{T.~Aushev}\affiliation{Institute for Theoretical and Experimental Physics, Moscow} 
  \author{A.~M.~Bakich}\affiliation{School of Physics, University of Sydney, Sydney, New South Wales 2006} 
  \author{B.~Bhuyan}\affiliation{Indian Institute of Technology Guwahati, Guwahati} 
% ...
\collaboration{The Belle Collaboration}
\noaffiliation


\begin{abstract}
We report a search for $B^{0}$ decays into invisible final states using a data sample
 of $657 \times 10^{6}$ $B\overline{B}$ pairs collected at the $\Upsilon(4S)$ resonance 
with the Belle detector at the KEKB $e^{+}e^{-}$ collider. The signal is 
identified by fully reconstructing a hadronic decay of the accompanying $B$ meson and 
requiring no other particles in the event. No significant signal is observed, 
and we obtain an upper 
limit of $1.3 \times 10^{-4}$ at the $90\%$ confidence level for the branching fraction of invisible $B^{0}$ decay.

\end{abstract}

\maketitle
\begin{figure}[htb]
\centering
\includegraphics[height=4.8cm]{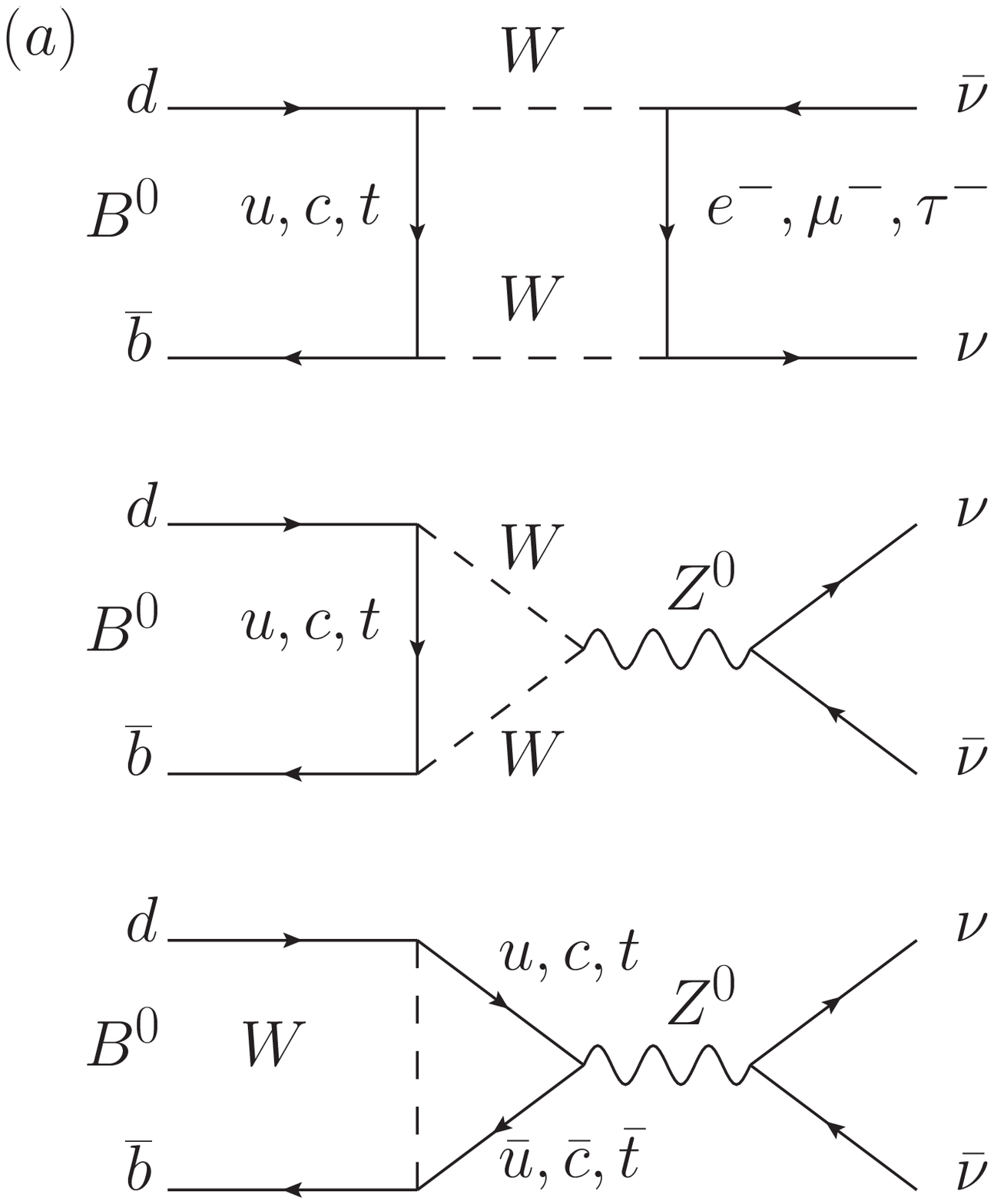}
\includegraphics[height=4.8cm]{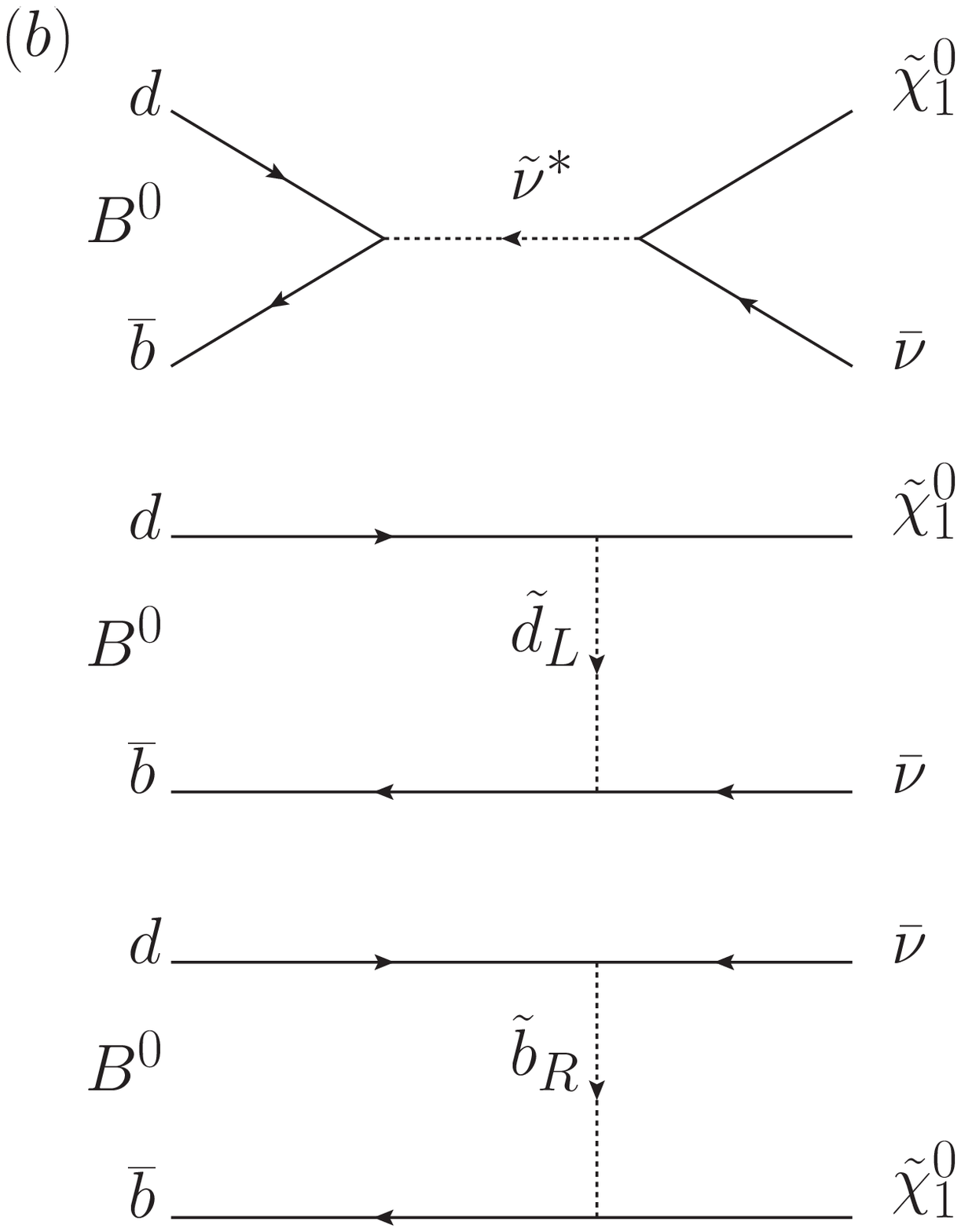} \\
\caption{Feynman diagrams for the SM process via $B^0 \to \nu \overline{\nu}$ (a)
and for new physics via $B^0 \to \tilde{\chi}^0_1 \overline{\nu}$ (b)\cite{nutev}.}
\label{fig:fenyman} 
\end{figure}

In the standard model (SM), the decay $B^0 \to \nu {\overline \nu}$ proceeds through the three annihilation diagrams shown in Fig.~\ref{fig:fenyman}(a). This decay  
is highly helicity suppressed with an expected branching fraction 
at the $10^{-20}$ level \cite{buchalla}. Because neutrinos participate only in weak interactions, the experimental signature is missing energy and momentum
corresponding to the presence of a $B^0$ meson in the event. 
New particles hypothesized by physics beyond the SM, such as $R$-parity violating supersymmetry,
can be involved in these $B$ decays, resulting in a final state with only weakly interacting particles and providing the same signature as in 
$B^0\to \nu {\overline \nu}$. For 
 instance, Ref.~\cite{nutev} discusses the $B$ decay into a neutrino and a 
 neutralino ($\tilde{\chi}^0_1$), shown in Fig.~\ref{fig:fenyman}(b); the branching fraction could be as high as $10^{-6}
- 10^{-7}$. Therefore, signals of invisible $B$ decays in current $B$ factory data would indicate new physics. So far no such signals were 
observed.
The first experimental result was provided by the BaBar 
Collaboration, with
$\br(B\to \mathrm{invisible}) < 2.2 \times 10^{-4}$ at the 90\% confidence level~\cite{Aubert:2004xy} with a semileptonic tagging method; recently, the upper limit was pushed to $2.4 \times 10^{-5}$ with more data and improved tagging efficiency by BaBar~\cite{babarnunubar:2012}.

In this paper we report the result of a search for $B$ decays to an invisible final 
state based on the data 
collected with the Belle detector at the KEKB asymmetric-energy (3.5 on 8 GeV) 
$e^+e^-$ collider~\cite{kur}.       
The data sample consists of $657 \times 10^{6}$ $B\overline{B}$ pairs accumulated at the $\Upsilon(4S)$ resonance, corresponding to an integrated luminosity of $606$ $\rm fb^{-1}$, and an additional 
$68~{\rm fb}^{-1}$ of off-resonance data recorded at a center-of-mass (CM) energy about $60$ MeV below the $\Upsilon(4S)$ resonance.
The Belle detector consists of a four-layer silicon vertex detector, a 50-layer central drift chamber (CDC), time-of-flight scintillation counters (TOF), an array of aerogel threshold Cherenkov counters (ACC), and a CsI(Tl)  electromagnetic calorimeter(ECL) located inside a superconducting solenoid coil that provides a 1.5~T magnetic field.
Outside the coil, the $K_L^0$ and muon detector (KLM), composed of resistive plate counters, detects $K_L^0$ mesons and identifies muons. The detector is described in detail elsewhere~\cite{aba}.
A GEANT3-based~\cite{geant} Monte Carlo (MC) simulation of the Belle detector is used to optimize the event selection and to estimate the signal efficiency.

Since the $\Upsilon(4S)$ decays to $B\overline{B}$ pairs, invisible $B$ decay candidates are identified by fully reconstructing a $B$ meson 
($\btag$) following the procedure of Ref.~\cite{ikado:frec} in hadronic modes,
  and then examining whether there are any other particles in the event. The neutral $\btag$ candidates are reconstructed through $B^0\to
D^{(*)-} h^+ $ decays, where $h^+$ denotes $\pi^+, \rho^+, a^+_1,$ or 
 $D_s^{(*)+}$~\cite{conjugate}.
Candidate $D^{*}_{(s)}$ mesons are identified through the channels $D^{*+}_s \to 
D_s^+ \gamma$ and $D^{*-}\to \overline{D}^0\pi^-$. Candidate $D_{(s)}$ mesons are reconstructed
using the following final states: $K^-\pi^+\pi^+$, $K^-\pi^+\pi^+\pi^0$, 
$K^+K^-\pi^+,$ $\ks\pi^+, \ks\pi^+\pi^0,$ and $\ks\pi^+\pi^+\pi^-$ for $D^+$;
$K^-\pi^+$, $\ks\pi^0, K^+K^-$, $K^-\pi^+\pi^0,$ $\ks\pi^+\pi^-,$ $K^-\pi^+\pi^+\pi^-$, 
and $\ks\pi^+\pi^-\pi^0$ for $D^0$; and $\ks K^+, K^+\pi^-\pi^+$, and $K^+K^-\pi^+$ for $D_s^+$. 

 Charged kaons and pions are identified using specific ionization from the CDC, time-of-flight information from the TOF, and Cherenkov light yield in the ACC.
 This information is combined to form a $K$-$\pi$ likelihood ratio 
$\mathcal{R}_{K/\pi} = \mathcal{L}_K/(\mathcal{L}_K+\mathcal{L}_\pi)$, 
where $\mathcal{L}_{K}$ $(\mathcal{L}_{\pi})$ is the likelihood that the track 
is a kaon (pion). Tracks with $\mathcal{R}_{K/\pi} > 0.6$ are regarded as kaons and $\mathcal{R}_{K/\pi} < 0.4$ as pions.
The typical selection efficiency for a 1.0 GeV/$c$ kaon (pion) 
is $83\%$ ($90\%$) while the misidentification probability for 1.0 GeV/$c$ kaons (pions) as pions (kaons) is around $6\%$ ($12\%$). Neutral $\ks \to \pi^+ \pi^-$ candidates are identified by pairing 
two opposite-sign charged tracks, both treated as pions, and then requiring that this pair have an 
invariant mass near the nominal $\ks$ mass with a vertex displaced from 
the $e^+e^-$ interaction point.
Candidate $K^0_L$'s are selected from KLM hit patterns that are not associated with any charged track~\cite{klcite}.
 Neutral pions are identified using the 
$\pi^0\to\gamma\gamma$ decay and requiring each photon to have a minimum energy
of 50 MeV and $\gamma \gamma$ mass between  
 0.115 GeV/$c^2$ and 0.156 GeV/$c^2$. The $\rho^+$ and $a_1^+$ meson candidates are reconstructed using the $\rho^+\to \pi^+\pi^0$ and $a_1^+\to \pi^+\pi^-\pi^+$ 
channels. 

\begin{figure}[ht]
\centering
\includegraphics[width=9.0cm]{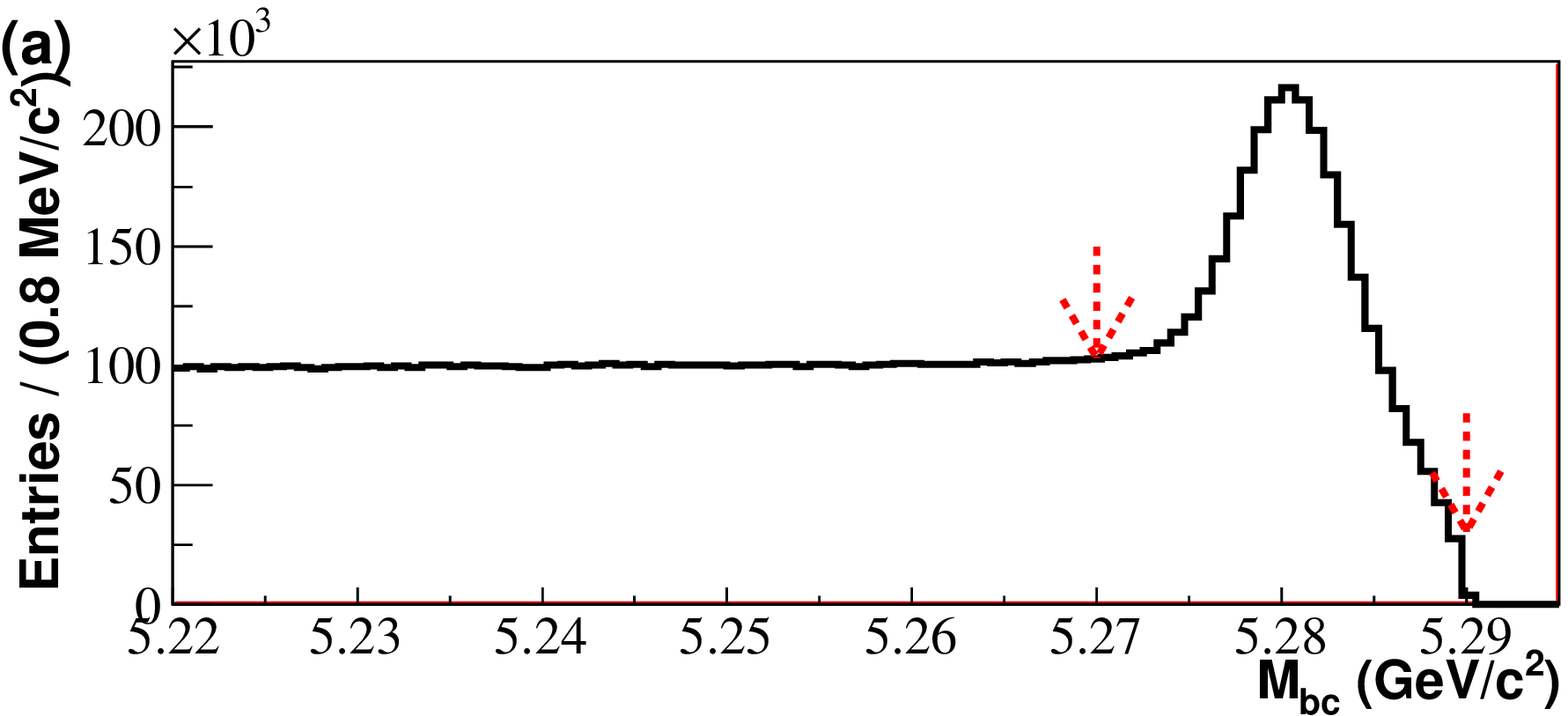} \\
\includegraphics[width=9.0cm]{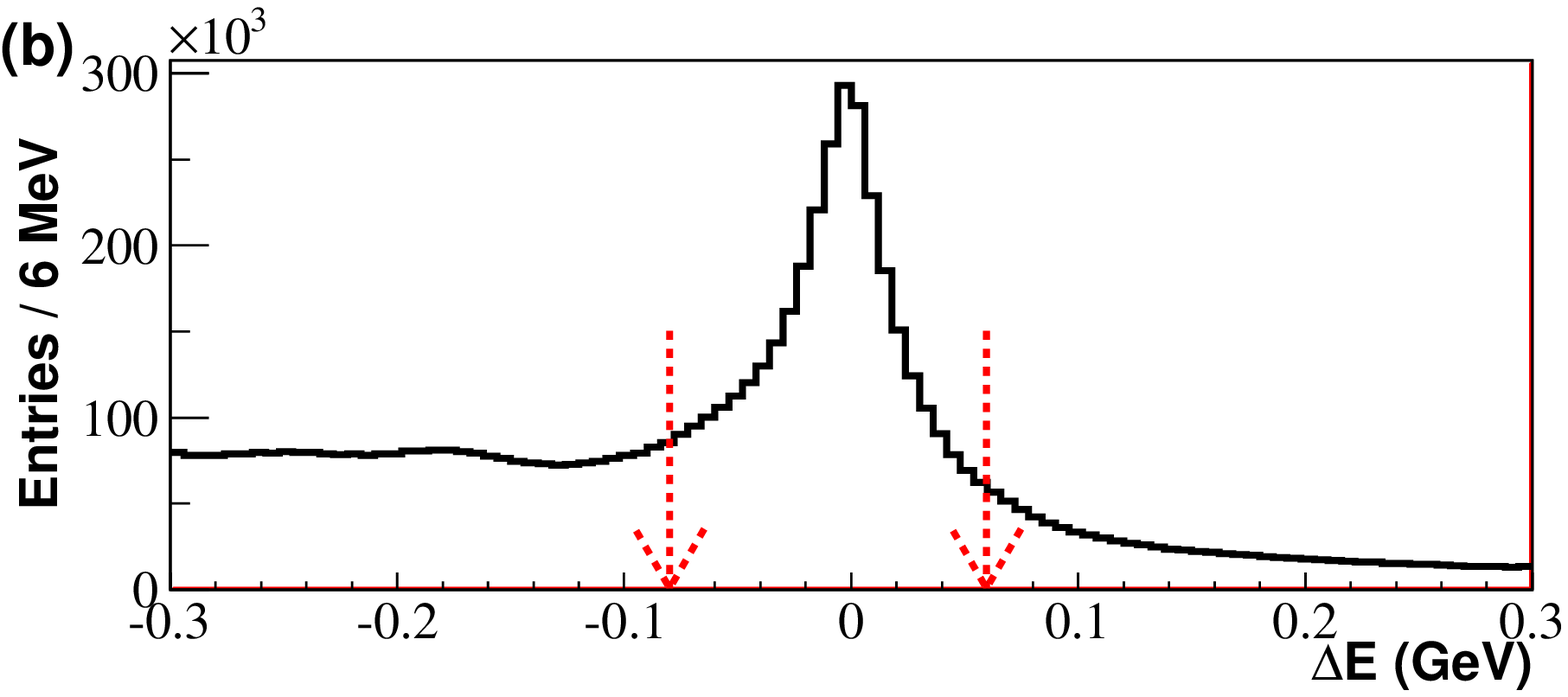} \\
\caption{The $\mb$ (a) and $\de$ (b) distributions for the $\btag$ candidates.
 Candidates having $\mb$ and $\de$ within the signal regions between the two arrows are used
 to search for $B$ decays to invisible final states.}
\label{fig:mbde}
\end{figure}

The selection of $\btag$ candidates is based on two kinematic variables: the
beam-energy constrained mass $\mb \equiv \sqrt{E^{2}_{\rm beam} - p_{B}^2}$ and the energy 
difference $\de \equiv E_B - \ebeam$, where $E_{B}$ and $p_{B}$ are the 
reconstructed energy and momentum of the $\btag$ candidate in the 
$e^+e^-$ CM frame, and $\ebeam$ is the beam-energy in 
this frame. The $\btag$ candidates are required to have $\mb>5.22\,{\rm GeV}/c^2$ and $|\de|<0.3\,{\rm GeV}$. Within this region, we define the signal 
region: $5.27$ GeV/$c^2$ $< \mb <5.29$ GeV/$c^2$ and $-0.08$ GeV
$< \de < 0.06$ GeV. Figure~\ref{fig:mbde} shows the $\mb$ and $\de$ distributions
of the $\btag$ candidates in data.   
If there are multiple $\btag$ candidates in an event, the 
candidate with the smallest $\chi^2$ is retained, where $\chi^2$ is computed
 using $\de$, 
the $D$ meson mass, and the mass difference between the $D^{*}$ and  
$D$ (for candidates with a $D^*$ in the final state), weighted using their expected
resolutions. We reconstruct $9.5 \times 10^{5}$ neutral $\btag$ candidates in total.
After identifying the $\btag$ candidate, we require no additional charged tracks nor $\pi^0$ or $K^0_L$
candidates in the rest of the event.

\begin{figure}[htp]
\centering
\includegraphics[width=4.2cm]{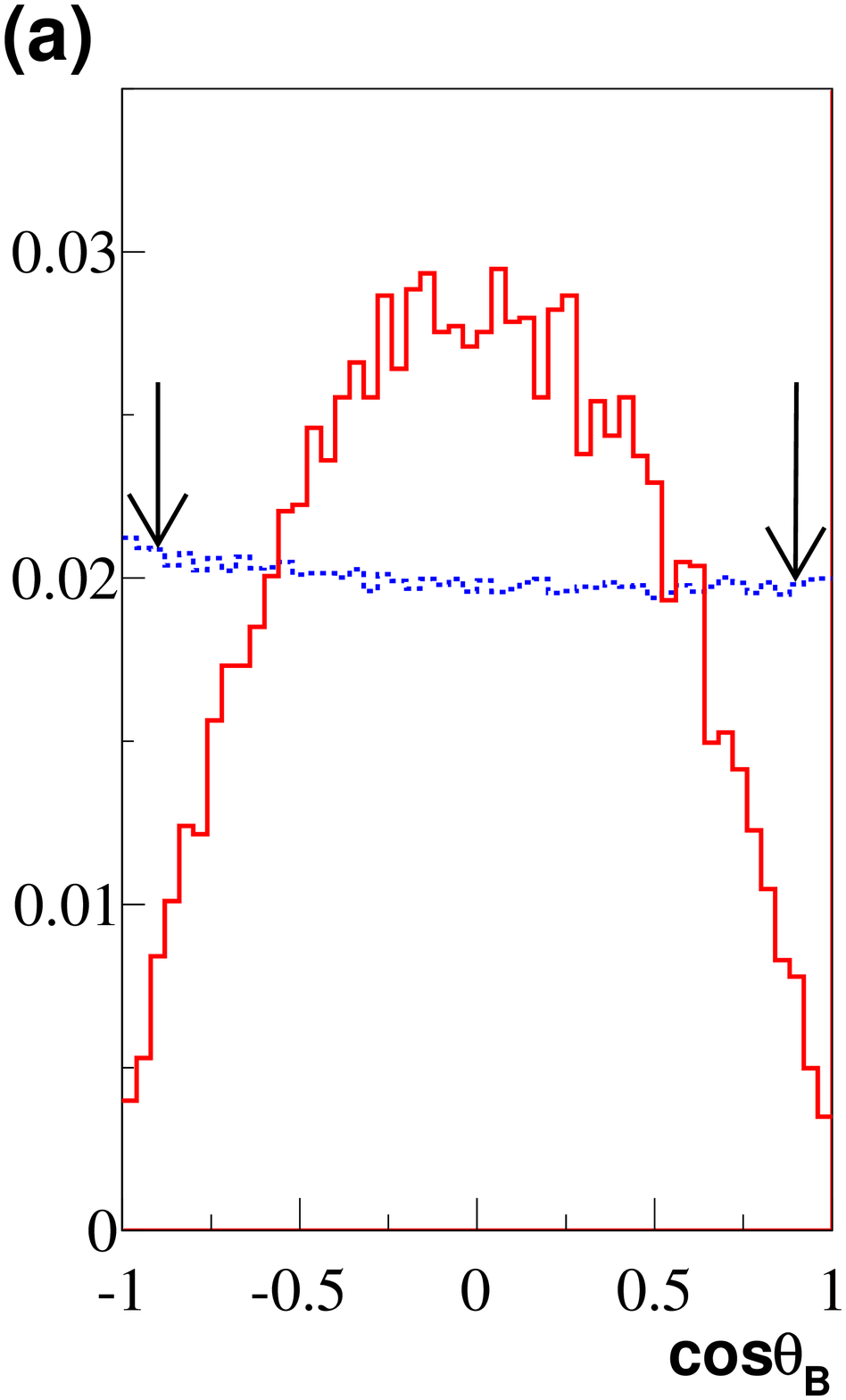}
\includegraphics[width=4.2cm]{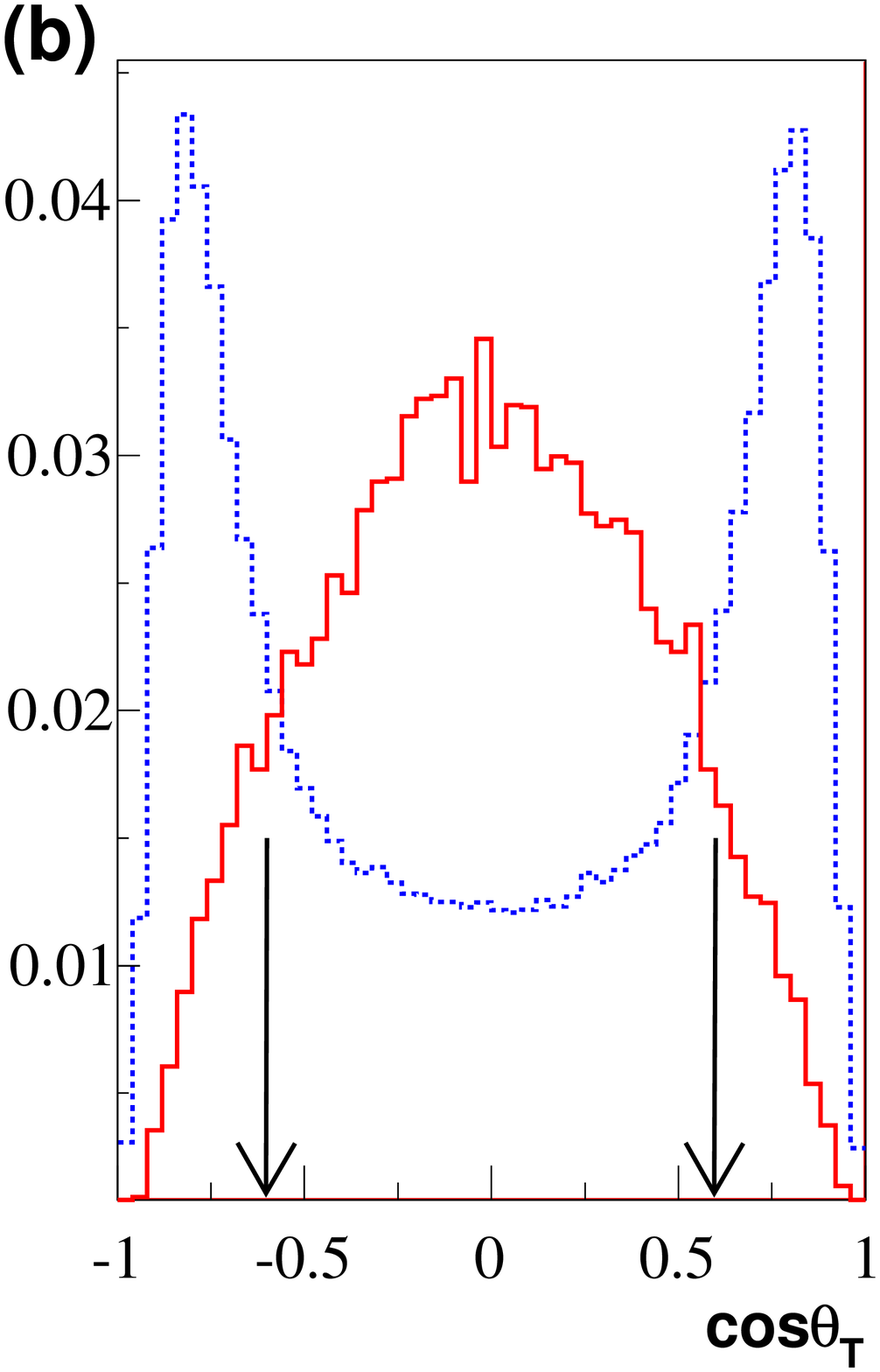}
\caption{\label{fig:cost}Normalized distributions of (a) $\cosb$ and (b) $\cost$ for the signal and continuum backgrounds. The solid histogram is the signal and the dashed histogram is the continuum background.}
\end{figure}

The dominant backgrounds are from $e^+e^- \to q\overline q ~( q=u,d,s,c )$ 
continuum events and $\bb$ decays with a $b\to c$ transition (generic $B$ background). 
Two variables are used to distinguish the signal and continuum events:
$\cosb$, defined as the cosine of the angle between the $\btag$ flight 
direction and the beam axis in the CM frame, and $\cost$, the cosine of the angle of the $\btag$ thrust axis with respect to the beam
axis in the CM frame. Clear differences in the distribution of each variable between signal 
and continuum background are shown in Fig.~\ref{fig:cost}, using the MC 
simulation.
We define the fit region as $-0.9 <\cosb<0.9$ and $-0.6<\cost<
0.6$.
The variable $\cosb$ is used in the fit to extract the signal 
yield.
Other backgrounds, such as rare $B$ decays via $b\to q ~(q = u, d, s)$ processes 
and $e^+e^- \to \tau^+\tau^-$ transitions, are also considered in the signal 
extraction and studied using large MC samples. The $\tau^+\tau^-$ background is  small and has an event topology similar to the continuum; therefore, 
 the continuum and $\tau^+\tau^-$ backgrounds are combined and called the 
non-$B$ background.

The most powerful variable to identify $B$ decays into the invisible final 
state is the residual energy in the ECL, denoted $\ecl$, which is 
the sum of the energies of ECL clusters that are not associated with the $\btag$ 
daughters. To further suppress the background, minimum energy thresholds
are required for clusters located in various ECL regions: 
50 MeV for the barrel ($32.2^\circ < \theta < 128.7^\circ$), 100 MeV for the forward endcap ($\theta < 32.2^\circ$), and 150 MeV for the backward endcap ($\theta > 128.7^\circ$).

The signal yield for invisible $B$ decays is extracted from an extended unbinned maximum likelihood fit to the $\ecl$ and $\cosb$ distributions. The likelihood is 
\begin{equation}
\calL= 
\frac{e^{-\sum_{j} n_j}}{N!}
\prod^{N}_{i=1}\left(\sum_{j} n_j \calP^i_j(\ecl,\cosb) \right),
\end{equation}
where $i$ is the event identifier; $n_j$ is the yield for 
category $j$, which corresponds to either signal, generic $B$, rare $B$ or non-$B$ 
background; and $\calP_j(\ecl,\cosb)$ is the product of the probability density 
functions (PDFs) $\calP(\ecl)$ and $\calP(\cosb)$, since we have verified that $\ecl$ and $\cosb$ are uncorrelated for each component. 
For each category, the $\ecl$ PDF is modeled as a histogram function, while the $\cosb$ PDF is described by a first or second 
order Legendre polynomial. The non-$B$ $\ecl$ PDF is constructed from off-resonance data, while all other PDFs are obtained using MC simulations.
The normalization of the rare $B$ background category is estimated from the MC simulation and is fixed 
in the fit.

The $\ecl$ simulation is validated using doubly tagged 
events in which the $\btag$ is fully reconstructed as described above and the
other $B$ is identified as $B^0 \to D^{(*)-} \ell^{+} \nu$ ($\ell = e, \mu$). Candidate $D^{*-}$ mesons are reconstructed via
$D^{*-}\to \overline{D}^{0}\pi^{-}$, followed by $\overline{D}^{0}\to K^+ \pi^-$, while 
$D^-$ is identified as  $D^{-} \to \ks \pi^{-}$ and $K^{+}\pi^{-}\pi^{-}$.
The track and $\pi^0$selections are applied here. Background contributions in the doubly tagged sample are found to be negligible; therefore, only loose selections on $D$ and $D^*$ masses and the mass squared of the undetected particles $m_{\rm miss}^2 = |{\mathbf P_{\rm beam}- \mathbf P_{\btag}-\mathbf P_{D^{(*)-}\ell^+}} |^{2}$ (where $\mathbf P$ denotes the four-momentum of the $e^+e^-$ system, $\btag$, or the $D^{(*)-}\ell^+$ system) are applied.

\begin{figure}[htpb]
\centering
\includegraphics[width=0.49\textwidth]{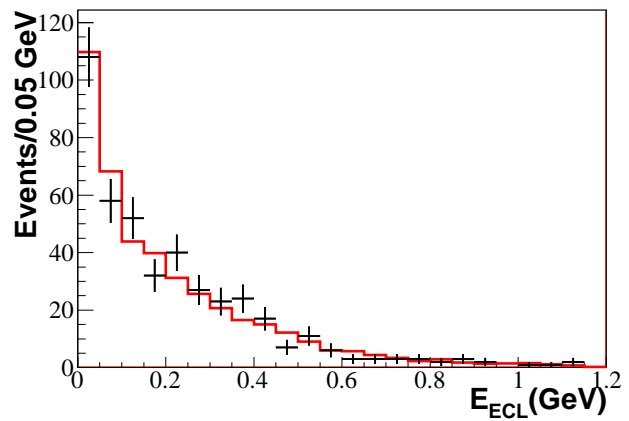}
\caption{$\ecl$ distribution for doubly tagged events, in which one $B$ is 
fully reconstructed and the other $B$ is reconstructed as 
$B^{0} \to D^{(*)-}\ell^{+}\nu$. Points with error bars are data, and the solid 
histogram is a signal MC simulation.
}
\label{fig:control}
\end{figure}

The observed $\ecl$ distributions for doubly tagged events, shown in Fig.~\ref{fig:control}, are found to 
be in good agreement with MC simulations.
The signal yields for control modes are obtained by fitting the $\ecl$ spectra while the efficiencies are estimated from MC samples.
The measured branching fractions with their errors, listed in Table~\ref{tab::control}, agree well with the Particle Data Group (PDG) values~\cite{pdg2010}. 
The $B^0 \to D^{(*)-} \ell^{+} \nu$ decays are also used to study the systematic uncertainty arising due to the 
track, $\pi^{0}$, and $K_{L}^{0}$ rejections as well as to calibrate the signal 
efficiency.
The aforementioned systematic uncertainties are estimated by comparing the efficiency before and after the application of those vetoes on data and MC. The data-MC efficiency ratios for track, $\pi^0$, and $K_{L}^0$ vetoes are $0.996 \pm 0.012$, $0.913 \pm 0.020$, and $1.096 \pm 0.020$, respectively. The central values are used to correct the MC efficiencies, while the statistical error is treated as a contribution to the systematic uncertainty. Since the central value of the track veto inefficiency is small, no scaling factor is applied on the veto efficiency. Instead, the sum of the inefficiency and the statistical error is quoted as a systematic uncertainty.

\begin{table}[htpb]
\caption{\label{tab::control}Summary of the fit result for $B^{0} \to D^{(*)-} \ell^{+} \nu$ samples (branching fractions in units of $10^{-3}$). The second and third columns show the products of branching fractions, where the error on the second column is statistical only.}
\begin{center}
\begin{tabular}{l|c|c}
\hline\hline
Mode & Measured result & PDG value\cite{pdg2010} \\
\hline
$B^{0} \to D^{*-} \mu^{+} \nu$ &  $1.41 \pm 0.20$ & $1.34 \pm 0.06$ \\
$B^{0} \to D^{*-} e^{+} \nu$ &  $1.62 \pm 0.18$ & $1.34 \pm 0.06$ \\
$B^{0} \to D^{-}(K\pi\pi) \mu^{+} \nu$ &  $1.99 \pm 0.21$ & $1.98 \pm 0.12$ \\
$B^{0} \to D^{-}(K\pi\pi) e^{+} \nu$ &  $1.93 \pm 0.14$ & $1.98 \pm 0.12$ \\
$B^{0} \to D^{-}(\ks\pi) \mu^{+} \nu$ &  $0.19 \pm 0.06$ & $0.22 \pm 0.02$ \\
$B^{0} \to D^{-}(\ks\pi) e^{+} \nu$ &  $0.21 \pm 0.05$ & $0.22 \pm 0.02$ \\
\hline\hline
\end{tabular}
\end{center}
\end{table}

\begin{figure}[htp]
\centering
\includegraphics[width=0.49\textwidth]{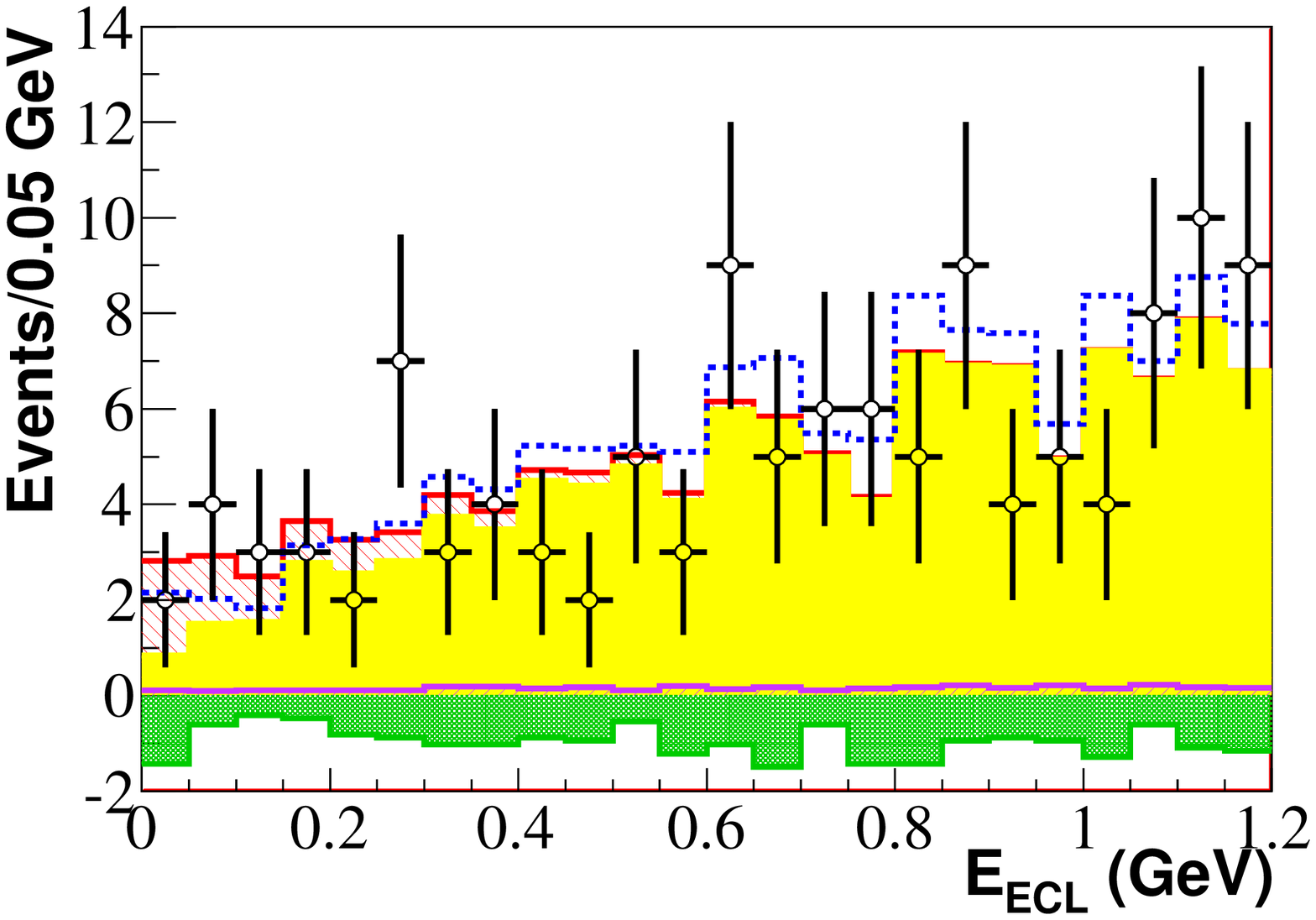}\\
\includegraphics[width=0.49\textwidth]{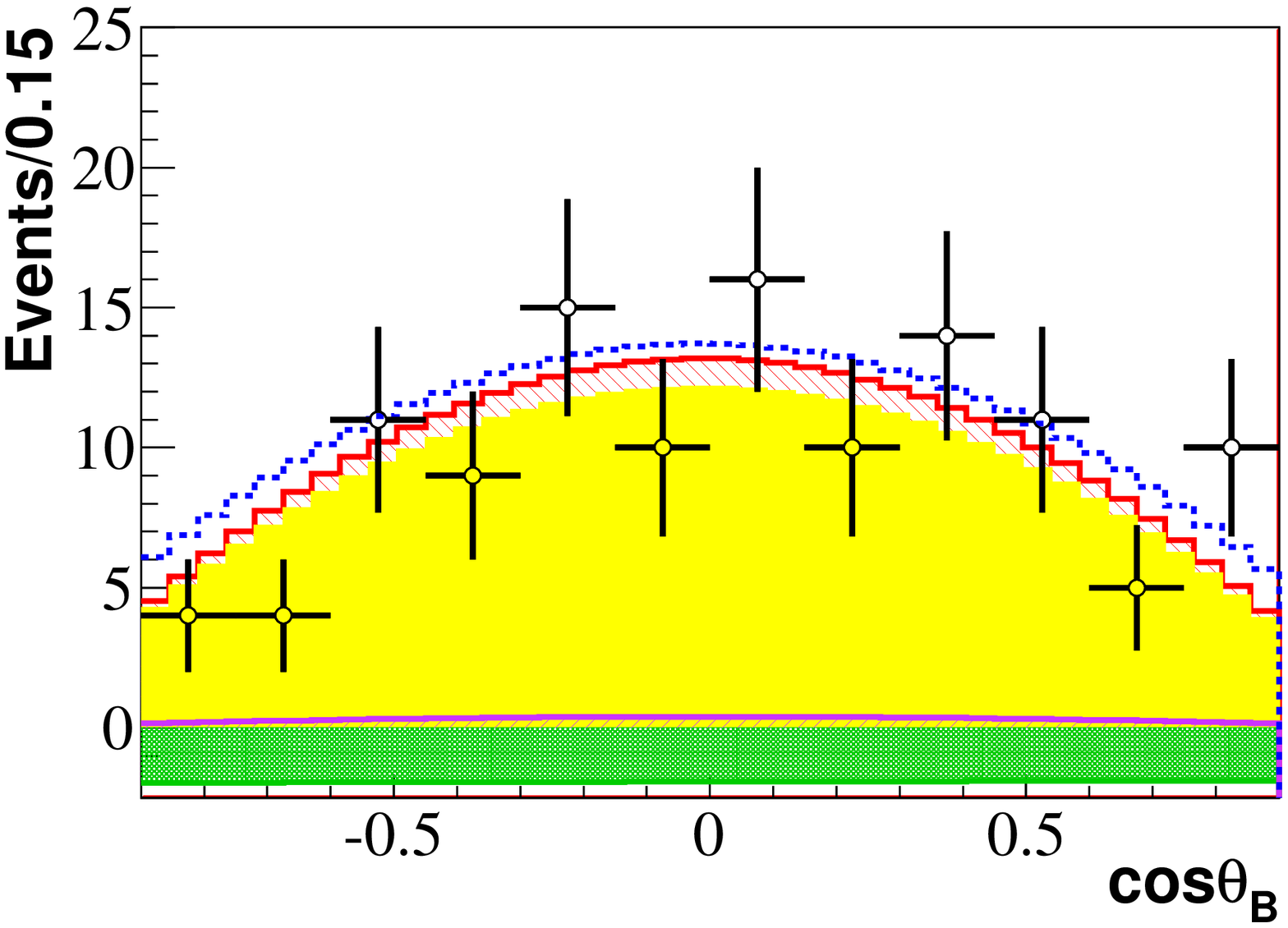}
\caption{\label{fig::open-box}The $\ecl$~(top) and $\cosb$~(bottom) 
distributions with fit results superimposed. Points with error bars are data. 
The red cross-hatched region is the signal component on the top of the total background shown in the yellow filled histogram. 
The blue dashed curve is the generic $B$ contribution, which is larger than the total because of the negative fit result for the non-$B$ background shown in the green dotted histogram. The purple hatched area corresponds to the rare $B$ contribution.}
\end{figure}

Table~\ref{tab::result} lists the signal and background yields for invisible $B$ decays from the fit 
while Fig.~\ref{fig::open-box} shows the $\ecl$ and $\cosb$ distributions
superimposed with the fit result. No significant signal is observed.
The signal efficiency, determined with MC simulations and later calibrated using 
the doubly tagged $B^0\to D^{(*)-} \ell^+\nu$ sample, is $(2.2 \pm 0.2) \times 10^{-4}$, where the error is dominated by the systematic uncertainty.

\begin{table}[htp]
\caption{\label{tab::result}Summary of fit yields for the signal 
and background. The normalization of the rare $B$ background contribution is fixed in the fit.}
\begin{center}
\begin{tabular}{l|c}
\hline\hline
Component & Yield \\
\hline
Signal			& $8.9^{+6.3}_{-5.5}$ \\
Generic $B$ background	& $131.6^{+21.9}_{-22.8}$ \\
Non-$B$ background	& $-23.2^{+21.6}_{-17.0}$ \\
Rare $B$ background	& $3.7$ \\
\hline
Observed events & $121$ \\
\hline\hline
\end{tabular}
\end{center}
\end{table}

The systematic uncertainty associated with the signal efficiency is 
dominated by the $\btag$ reconstruction efficiency. 
 The uncertainty on $\btag$ reconstruction is estimated by comparing the 
yield difference between data and the corresponding MC sample, generated with a 
proper mixture of generic $B$ and continuum events. The $\btag$ yields are extracted by fitting the $\mb$ distributions, and an uncertainty of $8.3\%$ is assigned. 
Systematic uncertainties arising from the requirement of no 
additional charged tracks nor $\pi^0$ and $K^0_L$ candidates are estimated 
to be $1.6\%$, $2.0\%$, and $2.0\%$, respectively, using $B^0 \to D^{(*)-} \ell^{+} \nu$ decays in data.
The uncertainty in the number of $B\overline{B}$ pairs is 
$1.4\%$.

\begin{table}[htp]
\caption{\label{tab::sys-pdf}Summary of systematic uncertainties arising from PDF modeling and components with fixed normalizations.}
\begin{center}
\begin{tabular}{l|c}
\hline\hline
Source & Events \\
\hline
Signal PDF			& Negligible \\
Generic $B$ PDF	& ${+1.6}/{-1.4}$ \\
Rare $B$ PDF 	& $\pm 0.1$ \\
Rare $B$ fixed yield & ${+0.2}/{-0.1}$ \\
Non-$B$ PDF	& ${+1.9}/{-1.3}$ \\
Binning effect & ${+1.7}/{-1.8}$ \\
\hline
Sum				& ${+3.0}/{-2.6}$ \\
\hline\hline
\end{tabular}
\end{center}
\end{table}

The uncertainties in the signal yield extraction are summarized in 
Table~\ref{tab::sys-pdf}.
The uncertainty due to fixing the normalization of the rare $B$ component is obtained by varying the rare $B$ yield by the estimated uncertainty ($\pm 1.9$ events). The corresponding variation in the signal
 yield, $^{+0.2}_{-0.1}$, is assigned as the systematic uncertainty. 
For each $\ecl$ PDF, we successively vary the content of each histogram bin by $\pm 1 \sigma$ to obtain a new PDF. The variation in the signal yield using the new PDF is calculated by performing an unbinned likelihood fit;
 the quadratic sum of all the variations gives the 
 systematic uncertainty for the PDF. The systematic uncertainty arising from $\cosb$ PDFs is negligible. Moreover, the effect of bin size
is also investigated by choosing different bin sizes to model the PDFs.
Again, the variation in the signal yield is considered as a systematic uncertainty.
The total systematic uncertainty is computed by summing all contributions listed in Table~\ref{tab::sys-pdf} in quadrature.

Since there is no significant signal observed, an upper limit at 90\%
confidence level (C.L.) is computed using the fit likelihood as a function of 
the branching fraction. The branching fraction is obtained from the signal yield from the fit, the signal selection efficiency, and the number of $B\overline{B}$ pairs. The likelihood at each branching fraction is obtained using 
Eq.~1 except that the signal yield is fixed in the fit. The systematic  
uncertainty of the measurement is taken into account by convolving the likelihood
function with a Gaussian whose width equals the systematic uncertainty ($\Delta \br$),

\begin{equation}
\calL_{\rm{smear}}(\br)=\varint\mathcal{L}(\br^{'})
\frac{e^{-\frac{(\br-\br^{'})^2}{2\Delta \br^2}}}{\sqrt{2\pi \Delta \br}}d\br^{'} \mbox{.}
\end{equation}

The upper limit on the branching fraction is estimated by integrating the likelihood function from zero to the bound that gives $90\%$ of the total area. We obtain $\br(B \to \mathrm{invisible})< 1.3\times 10^{-4}$
at the 90\% C.L. The expected upper limit, estimated by applying the same method on the MC sample, is $1.1 \times 10^{-4}$.

In conclusion, we have performed a search for  
$B \to \mathrm{invisible}$ decay with a fully reconstructed $\btag$ on a data sample of $657\times 10^{6}$ $B\overline{B}$ pairs collected at the 
$\Upsilon(4S)$ resonance with the Belle detector. No significant signal is observed, and we set an upper limit of $1.3 \times 10^{-4}$ at the $90\%$ confidence level for the branching fraction of invisible $B$ decay.
 The limit obtained for $B^0 \to \mathrm{invisible}$ decay is the most stringent constraint to date with a hadronic tagging method.

We thank the KEKB group for excellent operation of the
accelerator; the KEK cryogenics group for efficient solenoid
operations; and the KEK computer group, the NII, and 
PNNL/EMSL for valuable computing and SINET4 network support.  
We acknowledge support from MEXT, JSPS, and Nagoya's TLPRC (Japan);
ARC and DIISR (Australia); NSFC (China); MSMT (Czechia);
DST (India); INFN (Italy); MEST, NRF, BRL program with Grant No.~KRF-2011-0020333, GSDC of KISTI, and WCU (Korea);
MNiSW (Poland); MES and RFAAE (Russia); ARRS (Slovenia); 
SNSF (Switzerland); NSC and MOE (Taiwan); and DOE and NSF (USA).


\begin{thebibliography}{99}

\bibitem{buchalla} G. Buchalla and A. J. Buras, Nucl. Phys. B {\bf 400}, 225 (1993).
\bibitem{nutev}
A.~Dedes, H.~Dreiner, and P.~Richardson, Phys. Rev. D
{\bf 65}, 015001 (2001).
\bibitem{Aubert:2004xy} B. Aubert {\it et al.} (BaBar Collaboration), Phys. Rev. Lett. {\bf 93}, 091802 (2004).
\bibitem{babarnunubar:2012} J.~P. Lees {\it et al.} (BaBar Collaboration), arXiv:1206.2543 [hep-ex].
\bibitem{kur} S.~Kurokawa and E.~Kikutani, Nucl. Instr. Meth. A {\bf 499}, 1 (2003), and other papers included in this volume.
\bibitem{aba} A.~Abashian {\it et al.} (Belle Collaboration), Nucl. Instr. Meth. A {\bf 479}, 117 (2002).
\bibitem{geant} R.~Brun {\it et al.},
GEANT 3.21, CERN Report No. DD/EE/84-1 (1987).
\bibitem{ikado:frec} K. Ikado {\it et al.} (Belle Collaboration), Phys. Rev. Lett. {\bf 97}, 251802 (2006).
\bibitem{conjugate} The inclusion of charge-conjugate modes is implied throughout this paper.
\bibitem{klcite} K.~Abe {\it et al.} (Belle Collaboration), Phys. Rev. Lett. {\bf 87}, 091802 (2001); Phys. Rev. D {\bf 66}, 071102 (2002).
\bibitem{pdg2010} K. Nakamura {\it et al.} (Particle Data Group), J. Phys. G {\bf 37}, 075021 (2010) and 2011 partial update for the 2012 edition.
\end{thebibliography}
\end{document}